\documentclass[letterpaper, 10 pt, conference]{ieeeconf}
\IEEEoverridecommandlockouts                              
\usepackage{graphicx}

\usepackage[cmex10]{amsmath}
\usepackage{mathtools,amsthm}
\usepackage{times}
\usepackage{multicol}
\usepackage{amssymb, amsfonts}
\usepackage{cases}

\usepackage{enumerate}

\newcommand{\id}[1][3]{{I}}
\newcommand{\zero}[2]{{0}}
\newcommand{\zeros}[2]{{0}}

\newcommand{\multeqi}[2]{\begin{IEEEeqnarraybox}[][#2]{#1}}
\newcommand{\multeqf}{\end{IEEEeqnarraybox}}

\newcommand{\systemi}[1][rCL]{\left\lbrace\begin{IEEEeqnarraybox}[][c]{#1}}
\newcommand{\systemf}{\end{IEEEeqnarraybox}\right.}

\newcommand{\eqni}[1][rCL]{\begin{IEEEeqnarray}{#1}}
\newcommand{\eqnf}{\end{IEEEeqnarray}}

\newcommand{\nneqni}[1][rCL]{\begin{IEEEeqnarray*}{#1}}
\newcommand{\nneqnf}{\end{IEEEeqnarray*}}

\newcommand{\pmatrixi}{\begin{pmatrix}}
\newcommand{\pmatrixf}{\end{pmatrix}}

\newcommand{\bmatrixi}{\begin{bmatrix}}
\newcommand{\bmatrixf}{\end{bmatrix}}

\newcommand{\smatrixi}{\left[\begin{smallmatrix}}
\newcommand{\smatrixf}{\end{smallmatrix}\right]}

\newcommand{\enumi}{\begin{enumerate}}
\newcommand{\enumf}{\end{enumerate}}

\newcommand{\enumri}{\begin{enumerate}\renewcommand{\theenumi}{\textit{\roman{enumi}}}}
\newcommand{\enumrf}{\end{enumerate}}

\newcommand{\mytheorem}[2]{%
\newtheorem{t#2}{#1}%
\newenvironment{#2}{\begin{t#2}}{\end{t#2}}}

\theoremstyle{plain}
\newenvironment{remark}[1][Remark]{\begin{trivlist}
\item[\hskip \labelsep {\bfseries #1}]}{\end{trivlist}}

\mytheorem{Theorem}{theorem}
\mytheorem{Proposition}{proposition}
\mytheorem{Lemma}{lemma}
\mytheorem{Corollary}{corollary}
\mytheorem{Assumption}{assumption}
\mytheorem{Definition}{definition}
\mytheorem{Property}{property}

\newenvironment{system}[1][rCL]{\left\lbrace\begin{IEEEeqnarraybox}[][c]{#1}}{\end{IEEEeqnarraybox}\right.}

\usepackage{paralist}
\usepackage{textcomp}
\usepackage{gensymb}
\usepackage{color}

\graphicspath{figures/}

\newtheorem{hypothesis}{Assumption}

\usepackage[bookmarks=true]{hyperref}

\providecommand{\keywords}[1]{\textbf{\textit{Index terms---}} #1}

\hypersetup{pdfinfo={
   Author={Daniele Pucci},
   Title={On the  Existence of  Flight  Equilibria in Longitudinal Dynamics},
   CreationDate={D:20140625120000},
   Subject={Robotics},
   Keywords={Existence of Trim  Conditions, Longitudinal Dynamics},
}}

\begin{document}

\title{\LARGE \bf
On the  Existence of  Flight  Equilibria in Longitudinal Dynamics
}

\author{Daniele Pucci
\thanks{Dynamic Interaction Control  line, 
Istituto Italiano di Tecnologia, Via San Quirico 19 D, 16163 Genoa,
Italy
        {\tt\small daniele.pucci@iit.it}}%
}


\maketitle

\begin{abstract}
Any control law for aircraft asymptotic stabilization
requires the existence of an equilibrium condition, also called \emph{trim flight condition}. At a constant velocity flight,  for instance, there must exist an aircraft orientation such that
aerodynamic forces oppose the plane's thrust plus weight, and the torque balance equals zero. A closer look at the equations characterizing the trim conditions point out that the existence of aircraft equilibrium configurations cannot be in general claimed beforehand. By considering aircraft longitudinal linear dynamics, this paper shows that the existence of flight trim conditions is a  consequence of the vehicle shape or aerodynamics. 
These results are obtained independently from the aircraft flight envelope, and do not require any explicit expression of the aerodynamics acting on the vehicle. 
\end{abstract}


\section{Introduction}

The emergence of versatile flying robots  renewed the interest of the control community for flight control techniques.
Scale rotary-wing (e.g. quad-rotors) and fixed-wing (e.g. small airplanes) aerial robots represent, in fact, affordable platforms 
for testing modern control methods. One of the main challenges for these methods is the (usually) large flight envelope of aerial robots.  Fixed-wing flying robots, for instance, can often accomplish both vertical-take-off  and horizontal high-velocity cruising, thus  flying in very large flight envelopes. Hence, the  methods aimed at flight  stabilization of aerial robots inherit most of challenges that have been addressed separately for helicopters and airplanes, and require to work out general principles that little depend on the aircraft flight condition. This paper presents the first global results on the existence of aircraft  equilibrium configurations for any flight envelope and without the need of the specific aerodynamics acting on the vehicle.

\begin{figure}[tb]
\centering
\vspace*{0.4cm}
 \def\svgwidth{0.8\linewidth}
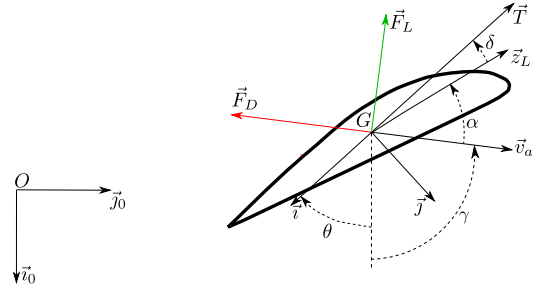
\caption{Propelled vehicle subject to aerodynamic forces.}
\label{fig:aerodyEff}
 \vspace{-1em}
\end{figure}

The application of Newton-Euler equations to the aircraft vertical plane yields  the following model, which has been extensively used for aircraft flight control and analysis in various scenarios and contexts  \cite[p. 166, p. 452]{2004_STENGEL}, \cite{2003_STEVENS,HUA_2013}:
\begin{IEEEeqnarray}{RCL}
	 \label{eq:dynamics0}
	 \IEEEyesnumber
	 m\vec{a} &=& \vec{F}_a(\vec{v},\theta,t) + m\vec{g} + \vec{T}(\theta), 
 	\IEEEyessubnumber 
 	\label{eq:newton2D} \\
	I_a \ddot{\theta} &=& M_{a} +M_{T},
	\label{eq:dyncooM} \IEEEyessubnumber
\end{IEEEeqnarray}
with $\vec{a}$ and $\ddot{\theta}$ the body linear and angualr acceleration, $\vec{v}$ and $\theta$ its velocity and orientation, $m$ and $I_a$ the aircraft mass and inertia, $\vec{F}_a$ and $M_a$ the  aerodynamic forces and moments,
and $\vec{T}$ and $M_T$ the force  and moment produced by vehicle's thrust -- see also  Figure~\ref{fig:aerodyEff}. Often, the  inputs are the thrust intensity $|\vec{T}|$ and some control surfaces influencing $M_a$.

At the equilibrium configuration, also called \emph{trim condition}, the control inputs make Eq.~\eqref{eq:dynamics0}   satisfied along the desired  trajectory.  For instance, at cruise horizontal flight, both left hand sides of Eq.~\eqref{eq:dynamics0}  equal  zero. Rendering these equations equal to zero, however, is in general far from obvious, all the more so because both the thrust and aerodynamic effects depend upon the vehicle orientation.  

The existence of  trim flight conditions is often deduced while seeking for the condition itself. A common procedure for doing so is to evaluate Eq.~\eqref{eq:dynamics0}  in specific flight envelopes. For instance, common assumptions are \emph{small} angles of attack \cite[p. 123]{2004_STENGEL} or \emph{small} aircraft velocities~\cite{HUA_2013}, which in turn yield either linear or negligible aerodynamics, respectively. Although numerical methods for finding  trim conditions exist~\cite{de_marco_general_2007,miller_aircraft_2003}, analytical characterizations of aircraft equilibrium existence irrespective from flight envelopes and aircraft shapes are still missing to the best of the author knowledge.

This paper complements our previous work~\cite{2012_PUCCI1,2013_PUCCI,phms15}  by investigating thoroughly the problem of the existence of aircraft trim conditions. In the same framework, we consider  aircraft longitudinal linear dynamics only, i.e. Eq.~\eqref{eq:newton2D}, so that we can work out general principles  independently of the aircraft torque actuation specifities.   
We here show that symmetric aircraft shapes  induce the existence of an equilibrium condition when the thrust force is parallel to the axis of symmetry. In addition, we show that bi-symmetric shapes  also ensure  the existence of a positive thrust force at the equilibrium independently of the thrust  direction with respect to the aircraft. These results are obtained independently from flight regime and envelope, and do not require any expression of the aerodynamics on the vehicle. For symmetric aircraft, 
we also show that the  trim condition existence  can be a  consequence of the \emph{aerodynamic stall}.

The paper is organized as follows. Section~\ref{sec:background} provides  notation and  background. Section~\ref{sec:problem-statement} states the addressed problem. Section~\ref{sec:existenceEquilibrium} presents the results on trim conditions existence.  Remarks and perspectives conclude the paper.

\section{Background}
\label{sec:background}

\subsection{Notation}
\label{sec:notation}

\begin{itemize}
\item  The $i_{th}$ component of  $x\in \mathbb{R}^n$ is denoted as $x_i$.
\item $\mathcal{I} = \{O;\vec{\imath}_0,\vec{\jmath}_0\}$ is an inertial frame with respect to (w.r.t.) which the vehicle's absolute pose is measured.
\item $\mathcal{B} = \{G;\vec{\imath},\vec{\jmath}\}$ is a frame attached to the body, and~$\vec{\imath}$ is parallel to the thrust  $\vec{T}$. This leaves two possible and opposite directions for $\vec{\imath}$. The direction chosen here
, i.e. $\vec{T} = -T\vec{\imath}$ with $T\in \mathbb{R}$,
is consistent with the convention used for VTOL vehicles. 
\item For the sake of brevity, $(x_1 \vec{\imath} + x_2 \vec{\jmath})$ is written as $(\vec{\imath},\vec{\jmath})x$.
\item $\{e_1,e_2\}$ is the canonical basis in $\mathbb{R}^2$,  $I$  the $(2 \times 2)$ identity matrix,
 $\vec{x} \cdot \vec{y}$  the scalar product between $\vec{x}, \vec{y}$. 
\item  Given a function of time $f: \mathbb{R} {\rightarrow} \mathbb{R}^n$,  its first time derivative is denoted
as $\frac{d}{dt}f = \dot{f}$. 
\item $G$ is the body's center of mass, 
 $\vec{p}:= \vec{OG} = (\vec{\imath}_0,\vec{\jmath}_0)x$ denotes the body's position.
$\vec{v} =\frac{d}{dt}\vec{p}= (\vec{\imath}_0,\vec{\jmath}_0)\dot{x}= (\vec{\imath},\vec{\jmath})v$ denotes the
the body's linear velocity, 
and $\vec{a}=\frac{d}{dt}\vec{v}$ the linear acceleration.
\item The vehicle orientation is given by the angle $\theta$  between $\vec{\imath}_0$ and $ \vec{\imath}$. 
The rotation matrix of  $\theta$ is $R(\theta)$. The column vectors of $R$ are the vectors of coordinates of
$\vec{\imath},\vec{\jmath}$  in $\mathcal{I}$. The matrix $S = R(\pi/2)$ is a unitary skew-symmetric matrix. 
The body's angular velocity is $\omega := \dot{\theta}$.

\end{itemize}

\subsection{Aerodynamic forces}
\label{par:aerodyForce}
\emph{Steady} aerodynamic forces at constant 
Reynolds and Mach numbers
can be written as follows~\cite[p. 34]{2010_AND}
\begin{IEEEeqnarray}{RLL}
	\label{aerodynamicForce2D}
	\vec{F}_a &=& k_a|\vec{v}_a|\left[c_L(\alpha)\vec{v}^\perp_a -c_D(\alpha)\vec{v}_a\right],
\end{IEEEeqnarray}
with $k_a {:=}\tfrac{\rho \Sigma}{2} $, $\rho$ the \emph{free stream} air density, 
$\Sigma$ the characteristic surface of the vehicle's body, 
$c_L(\cdot)$ the \emph{lift coefficient}, $c_D(\cdot)>0$  the \emph{drag coefficient} ($c_L$ and $c_D$ are called
\emph{aerodynamic characteristics}), $\vec{v}_a = \vec{v}-\vec{v}_w$ the \emph{air velocity},    $\vec{v}_w$  the wind's velocity,
$\vec{v}^\perp_a$ obtained by rotating the vector $\vec{v}_a$ by $90^\circ$ anticlockwise, i.e.  
	$\vec{v}^\perp_a = v_{a_1}\vec{\jmath} - v_{a_2}\vec{\imath}$,
and $\alpha$ the angle of attack. This latter variable is  
here 
defined as the angle between the
body-fixed {\em zero-lift} direction~$\vec{z}_L$, along which airspeed does not produce lift, 
and the airspeed $\vec{v}_a$, i.e. 
\begin{equation}
	\alpha := \text{angle}(\vec{v}_a,\vec{z}_L).
	\label{eq:angleOfAttacVec}
\end{equation}
The model~\eqref{aerodynamicForce2D} neglects the so-called \emph{unsteady aerodynamics}, e.g., the flow pattern effects induced by \emph{fast} angular velocity motions~\cite[p.199]{2004_STENGEL}. This assumption is commonly accepted in the robotics  literature dealing with large flight-envelope  control~\cite{naldi2011optimal,jung2012development,4803783,muraoka2009quad,frank2007hover,maqsood2010optimization}: in fact, building global, unsteady aerodynamic models for on-line control  purposes is still a  challenge even for the specialized aerodynamic literature~\cite{bruntonPhd,brunton2014state}.

Now, denote the constant angle between the zero-lift direction $\vec{z}_L$ and the thrust $\vec{T}$  as $\delta$, i.e. 
$\delta := \mathrm{angle}(\vec{z}_L,\vec{T}),$
and  the angle between 
the gravity $\vec{g} := g\vec{\imath}_0$ and~$\vec{v}_a$ as $\gamma$, i.e. 
$	\gamma := \mathrm{angle}(\vec{\imath}_0,\vec{v}_a).$ Then, one has
(see Figure~\ref{fig:aerodyEff}):
\begin{IEEEeqnarray}{RCL}
	\alpha  &=& \theta -\gamma  +(\pi -\delta)
	\label{eq:angleOfAttack}, \\
	\IEEEyesnumber
	\label{eq:VaComponents}
	v_{a_1} &=& -|\vec{v}_a|\cos(\alpha+\delta)  \quad \ \ 
	\label{eq:va1}  \IEEEyessubnumber  \\
	v_{a_2} &{=}&  \hspace{0.34cm} |\vec{v}_a|\sin(\alpha+\delta). \IEEEyessubnumber
	 \label{eq:va2}
\end{IEEEeqnarray}

\subsubsection{Symmetric shapes}
\label{sec:shape}
\begin{figure}[t]
  \centering
 \def\svgwidth{0.71\linewidth}
        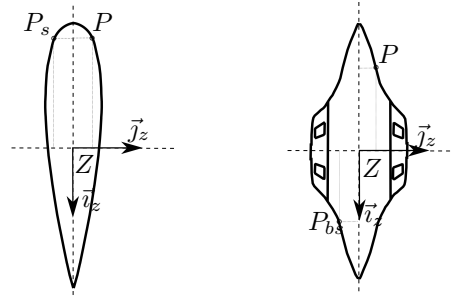
   \caption{Examples of symmetric and bisymmetric bodies.}
   \label{fig:symAndBisym}
   \vspace{-0.4cm}
\end{figure}
To characterize two kinds of  shape symmetries and their properties, let $\mathcal{B}_z~=~\{Z;\vec{\imath}_z,\vec{\jmath}_z\}$ be an orthonormal frame, and 
$P$ a point of the body surface~$\mathcal{S}$ -- see Figure~\ref{fig:symAndBisym}. 
Consider the vector~$\vec{ZP}$ and its expression w.r.t. the frame $\mathcal{B}_z$, i.e.
$\vec{ZP} :=  x\vec{\imath}_z+y\vec{\jmath}_z, \text{ with }  x,y \in~\mathbb{R}.$
Then, \emph{symmetric} and \emph{bisymmetric} shapes 
satisfy what follows.
\begin{hypothesis}[\textbf{Symmetry}]
\label{hy:symmetries2D} 
There exists a choice for the frame $\mathcal{B}_z$ such that the point $P_s$ defined by the vector 
$\vec{ZP}_s = x\vec{\imath}_z-y\vec{\jmath}_z$ 
belongs to $\mathcal{S}$ for any point $P$ of the  surface $\mathcal{S}$. Then, the  shape is said to be \emph{symmetric},
with axis of symmetry given by $\{Z,\vec{\imath}_z\}$.
\end{hypothesis}
\begin{hypothesis}[\textbf{Bisymmetry}]
\label{hy:bisymmetries2D}
There exists a choice for the frame $\mathcal{B}_z$ such that the point $P_{bs}$ defined by 
$\vec{ZP}_{bs} = -x\vec{\imath}_z-y\vec{\jmath}_z$ 
belongs to $\mathcal{S}$ for any point $P$ of the  surface $\mathcal{S}$. Then, the  shape is said to be \emph{bisymmetric},
with axes of symmetry given by $\{Z,\vec{\imath}_z\}$ and $\{Z,\vec{\jmath}_z\}$.
\end{hypothesis}
We assume that an axis of symmetry identifies two zero-lift-directions. Then, 
we choose the zero-lift-direction $\vec{z}_L$ in~\eqref{eq:angleOfAttacVec}  parallel to an axis of symmetry, which implies that 
$c_L(0)=c_L(\pi)=0$. Note that this choice still leaves two possible and opposite directions for 
the definition of the vector $\vec{z}_L$, which in turn may reflect
in two possible values
of the angle~$\delta$. 
Without loss of generality, the direction here chosen is that minimizing the angle~$\delta$. 

In light of the above,
a symmetric shape induces aerodynamic
 characteristics $c_D( \alpha)$ and $c_L( \alpha)$ that are even and odd functions, respectively. 
\begin{property}
\label{propertySymmetricShape}
If the body shape $\mathcal{S}$ is symmetric and the zero-lift-direction $\vec{z}_L$ 
is parallel to the axis of symmetry, then 
\begin{IEEEeqnarray}{RCL}
	\IEEEyesnumber
	    \label{LiftPropertiesSymmetricB}
 		c_D( \alpha) &=& c_D( -\alpha),   \quad \quad c_L( \alpha) = -c_L( -\alpha), \quad \forall \alpha, \IEEEyessubnumber \IEEEeqnarraynumspace \\ 
		c_L( 0) &=& c_L( \pi) = 0. \IEEEyessubnumber \label{cLZeroCLPiUZ} 
		\IEEEeqnarraynumspace
\end{IEEEeqnarray}
\end{property}
Bisymmetric shapes have an additional symmetry about the axis $\vec \jmath_z$, thus implying the invariance
of the aerodynamic forces w.r.t. body rotations of $\pm \pi$. Then, 
the aerodynamic characteristics of bisymmetric shapes are $\pi-$periodic functions versus the angle 
$\alpha$. 
\begin{property}
\label{propertyBisymmetricShape}
If the body shape $\mathcal{S}$ is bisymmetric and the zero-lift-direction $\vec{z}_L$ 
is parallel to an axis of symmetry, then the aerodynamic coefficients satisfy~\eqref{LiftPropertiesSymmetricB} and
\begin{IEEEeqnarray}{RCL}
	    \label{LiftPropertiesBSymmetricB}
		c_D( \alpha) &=& c_D( \alpha \pm \pi), \quad \quad
		c_L( \alpha) = c_L( \alpha \pm \pi), \quad \forall \alpha.  \IEEEeqnarraynumspace 
\end{IEEEeqnarray}
\end{property}

\section{Problem statement}
\label{sec:problem-statement}

Assume that the control objective is the asymptotic stabilization of  a reference velocity.
Let $\vec{v}_r(t)$ denote the differentiable reference velocity, and  $\vec{a}_r(t)$ its  time derivative, 
i.e. $\vec{a}_r(t)~=~\dot{\vec{v}}_r(t)$. Now,
define the velocity error as follows
\begin{IEEEeqnarray}{rcl}
  \label{velocityError}
  \vec{e}_v := \vec{v}-\vec{v}_r.
\end{IEEEeqnarray} 
Using System~\eqref{eq:dynamics0} one obtains the following error model
\begin{IEEEeqnarray}{rCL}
	\label{eq:errorsDynamics}
	m\dot{\vec{e}}_v &=& \vec{F}-T\vec{\imath},
	\IEEEyessubnumber \label{eq:dynamicsVelocityError} \\
	\dot{\theta} &=& \omega,
 	\IEEEyessubnumber \label{eq:omegaSExt}
\end{IEEEeqnarray}
with $\vec{F}$  the \emph{apparent external force} defined by
\begin{IEEEeqnarray}{RCL}
  \label{externalApparentForce}
  \vec{F} &:=& m\vec{g} +\vec{F}_a-m\vec{a}_r. 
\end{IEEEeqnarray}
 Eq.~\eqref{eq:dynamicsVelocityError} indicates that the  condition $\vec{e}_v \equiv 0$ requires
\begin{IEEEeqnarray}{RCL}
   T\vec{\imath}(\theta) = \vec{F}({\vec{v}_r(t),\theta,t}),  \forall t, 
   \label{equilibriumCondition}
\end{IEEEeqnarray}
which 
in turn 
implies  
\begin{IEEEeqnarray}{RCL}
\IEEEyesnumber
  \label{equilibiumConditions}
  T &=& \vec{F}({\vec{v}_r(t),\theta,t})\cdot \vec{\imath}(\theta), \IEEEyessubnumber \label{equilibiumThrustCondition} \\
  0 &=& \vec{F}({\vec{v}_r(t),\theta,t}) \cdot \vec{\jmath}(\theta) \quad \forall t.  \IEEEyessubnumber \label{equilibiumOrinetationCo}
\end{IEEEeqnarray}
The existence of an orientation $\theta$ such that Eq.~\eqref{equilibiumOrinetationCo} is satisfied cannot be ensured \emph{a priori}. In fact,
the apparent external force $\vec{F}$ depends on the vehicle's orientation, and any change of this orientation affects both vectors 
$\vec{F}$ and $\vec{\jmath}$.
The dependence of the apparent force $\vec{F}$ upon the  orientation~$\theta$ comes from the dependence of the aerodynamic force $\vec{F}_a$ 
upon $\alpha$ (see Eqs.~\eqref{aerodynamicForce2D} and~\eqref{eq:angleOfAttack}). 

Hence, in view of Eq.~\eqref{equilibiumOrinetationCo}, we  state the definition next.
\begin{definition}
 \label{def:eqOrie}
 An \emph{equilibrium orientation} $\theta_e(t)$ 
 is a time function such that Eq.~\eqref{equilibiumOrinetationCo} is satisfied with $\theta =\theta_e(t)$. 
\end{definition}

\noindent
The \emph{existence} of an equilibrium orientation is a necessary condition for the asymptotic stabilization of a reference velocity.
Note, however, that  a reference velocity $\vec{v}_r(t)$ may  induce several equilibrium orientations. 
To classify the \emph{number} of these  orientations, define the set $\Theta_{\vec{v}_r}(t)$ as
\begin{IEEEeqnarray}{RCL}
  \label{thetaSet}
 \Theta_{\vec{v}_r}(t) {:=} \bigg\{ \theta_e(t) {\in} \mathbb{S}^1 : 
 \vec{F}({\vec{v}_r(t),\theta_e(t),t}) \cdot \vec{\jmath}(\theta_e(t)) {=} 0 \bigg\}. \IEEEeqnarraynumspace
\end{IEEEeqnarray}

Remark that given an equilibrium orientation $\theta_e(t)$, the thrust intensity $T$ at the equilibrium configuration is 
given by Eq.~\eqref{equilibiumThrustCondition}  with $\theta = \theta_e(t)$. 
The existence of an equilibrium orientation ensuring a positive thrust is
of particular importance, since positive-thrust limitations represent a common constraint when considering aerial vehicle control and planning.
To characterize this existence, define
\begin{IEEEeqnarray}{RCL}
 \label{positiveThrustThe}
 \Theta^+_{\vec{v}_r}(t) {:=} \bigg\{ &\theta_e(t)& \in \Theta_{\vec{v}_r}(t) : 
 \vec{F}({\vec{v}_r,\theta_e,t})\cdot \vec{\imath}(\theta_e) \geq 0 \bigg\}.
  \IEEEeqnarraynumspace
\end{IEEEeqnarray}

\section{Existence of  equilibrium orientations}
\label{sec:existenceEquilibrium}
We know from experience that airplanes do fly. So, given a reference velocity,
the equilibrium orientation should exist in most cases. 
One may then conjecture that the existence of  equilibrium orientations follows from
aerodynamic properties that hold independently of the body's shape, 
alike the \emph{passivity}  of aerodynamic forces.
In particular, the (steady) aerodynamic force always resists the relative 
motion of the body, and one may believe that this general property induces the existence of flight equilibria. 
Next lemma, however, shows that the passivity of aerodynamic forces  is not a sufficient condition
for the existence of equilibrium orientations.
\begin{lemma}
    \label{lemma:dissipativityNotSufficient}
    The passivity of the aerodynamic force, i.e.
    \begin{IEEEeqnarray}{r}
	\vec{v}_a \cdot \vec{F}_a(\vec{v}_a,\alpha) \leq 0, \quad \forall (\vec{v}_a,\alpha), \IEEEeqnarraynumspace \label{dissipativity}
    \end{IEEEeqnarray}
    is not a sufficient condition for the existence of an equilibrium orientation.
\end{lemma}
The proof is given in the Appendix.
Another route that we may follow to conclude about the existence of an equilibrium orientation is  
considering specific classes 
of body's shapes. 
The  theorem next presents results on the equilibrium orientation existence by considering 
symmetric and bisymmetric shapes as defined in Section \ref{sec:shape}.
\begin{theorem} 
	\label{th:existence}
	Assume that the  aerodynamic coefficients $c_L(\alpha)$ and $c_D(\alpha)$ are continuous 
	functions, and that the reference velocity is differentiable, 
	i.e. $\vec{v}_r(t) \in \mathbf{\bar{C}}^1$. 
\begin{enumerate}
		\item[i)] If the body shape is symmetric 
		and the thrust is parallel to the its axis of symmetry, 
		then there exist at least two equilibrium orientations 
		for any reference velocity, i.e.
		 \[ \emph{\text{cardinality}}(\Theta_{\vec{v}_r}(t)) \geq 2 \quad \forall t, \quad \forall \vec{v}_r(t) \in \mathbf{\bar{C}}^1. \]
		\item[ii)] If the body's shape is bisymmetric,
		then there exists at least one equilibrium orientation  ensuring a 
		positive-semidefinite thrust  
		for any 
		reference velocity, i.e.
		\[\emph{\text{cardinality}}(\Theta^+_{\vec{v}_r}(t)) \geq 1 \quad \forall t, 
		\quad \forall \vec{v}_r(t) \in \mathbf{\bar{C}}^1, \]
		whatever the (constant) angle $\delta$
	  between the zero-lift direction and the thrust force.
	\end{enumerate}
\end{theorem}

The proof is given in the Appendix. Theorem~\ref{th:existence} points out that the existence of an equilibrium orientation follows from the symmetry properties 
of the body's shape, 
independently of its aerodynamics and specific families of reference velocities. 
More specifically, 
Item~$i)$ asserts that for symmetric body's shapes powered by a thrust force parallel to 
their axis of symmetry, e.g. $\delta=0$,
the existence of (at least) two distinct equilibrium orientations is guaranteed for any reference velocity. 
Item~$ii)$ 
states that the bisymmetry of the  shape implies the existence of an 
equilibrium orientation independently of the thrust direction with respect to 
the body,
i.e. the angle $\delta$. Of most importance, this item points out that 
the shape's bisymmetry implies the existence of an equilibrium orientation inducing  a positive-semidefinite thrust intensity independently of reference trajectories. 

Now, assume that the body's shape is symmetric and not bisymmetric. 
If the thrust force is not parallel to the shape's axis of symmetry, the assumptions of
Theorem~\ref{th:existence} are not satisfied and the existence of an equilibrium orientation cannot be asserted.
Yet,  common sense makes us think that an equilibrium orientation still exists.

By considering symmetric shapes, the next theorem states conditions ensuring the existence of an equilibrium orientation 
independently of reference velocities and thrust directions w.r.t. the body's zero-lift direction. 

\begin{figure}[t!]
  \centering
  \vspace*{-0.2cm}
  \def\svgwidth{0.6\linewidth}\input{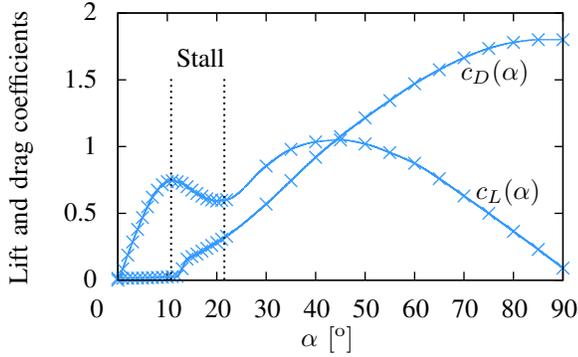}
  \caption{Aerodynamic coefficients of NACA 0021 (length $l = 0.91 m$, 
  chord  $c = 0.15m$,  
Reynolds and Mach numbers of $R_e = 16 \cdot 10^4$, $M = 0.3$).}
  \vspace*{-0.4cm}
  \label{fig:bifurcationsAndStall}
\end{figure} 

\begin{theorem}
	\label{th:existenceDeltaDifZeroSymm}
	Consider symmetric shapes. Assume that the aerodynamic coefficients $c_L(\alpha)$ and $c_D(\alpha)$ 
	are continuous functions, and that $c_D(\pi) > c_D(0)$. 
	If there exists an angle $\alpha_s \in (0,\pi/2)$ such that $c_L(\alpha_s) > 0$ and 
	  \begin{IEEEeqnarray}{RCL}
		  \label{conditionMuNe0}
		  \tan(\alpha_s) &\leq& \frac{c_D(\alpha_s)-c_D(\pi)}{c_L(\alpha_s)},
	  \end{IEEEeqnarray}
	  then there exists at least one equilibrium orientation 
	  for any 
	  reference velocity, i.e.
		\[ \emph{\text{cardinality}}(\Theta_{\vec{v}_r}(t)) \geq 1 \quad \forall t, 
		\quad \forall \vec{v}_r(t) \in \mathbf{\bar{C}}^1, \]
		whatever the (constant) angle $\delta$ between the zero-lift direction and the thrust force.
\end{theorem}
The proof is given in the Appendix. 
Theorem~\ref{th:existenceDeltaDifZeroSymm} requires some knowledge of the body's aerodynamic coefficients 
to assert the existence of the equilibrium orientation.
The key hypothesis in Theorem~\ref{th:existenceDeltaDifZeroSymm} is the existence of an angle $\alpha_s$ such that the
condition~\eqref{conditionMuNe0} is satisfied. 
Seeking for this angle requires some aerodynamic data, and it may be airfoil and flow regime dependent.
Recall, however, that  \emph{stall phenomena} 
(see, e.g., Figure~\ref{fig:bifurcationsAndStall}) involve rapid, usually important, lift decreases and drag increases.
Then the likelihood of satisfying the condition~\eqref{conditionMuNe0} 
with $\alpha_s$ belonging to the stall region is very high. We verified that Theorem~\ref{th:existenceDeltaDifZeroSymm} indeed applies 
with $\alpha_s$ belonging to the stall region for 
the NACA airfoils 0012, 0015, 0018, and 0021 at $M = 0.3$ and several Reynolds numbers (data taken from~\cite{CYBERIAD}).



\section*{Acknowledgments}
The author is infinitely thankful to Tarek Hamel, Pascal Morin, and Claude Samson for their support, guidance, and help during the
development of this research project.

\section{Conclusions}
This paper addressed the problem of finding  conditions ensuring flight equilibria in longitudinal dynamics, also called \emph{aircraft trim conditions}. The results presented here hold for any aircraft flight envelop and aerodynamics, and mainly depend on the geometric properties of the vehicle shape only. 

A major difficulty when seeking for the existence of trim conditions   comes from the dependence of aerodynamics upon the vehicle orientation -- see~\cite{pucciPhd} for  details. In fact, any change of this orientation induces a change of both the vehicle aerodynamics and thrust, which renders the possibility of satisfying the equations of motion along the desired trajectory not obvious. 
What we have shown is that aircraft trim conditions exist in most cases if the airfoil is 
quasi-symmetric, and this existence
is independent from the  flight envelope and  thrust direction relative to the body.

To work out general principles, the existence of trim conditions is here investigated only for linear longitudinal flight dynamics, so the specific torque actuation of the vehicles is not taken into account. In other words, we present here necessary conditions for the existence of trim conditions, which  should be then verified for the specific aircraft angular dynamics that depends on the vehicle actuation specifities.

Extensions of this work concern the analysis of the trim condition  for specific classes of vehicles. 
Another future direction is clearly the extension to  the three-dimensional case, and eventually to flying multi-body robots~\cite{7997895,8624985} that would  involve more complex aerodynamics effects.

\bibliographystyle{IEEEtran}
\bibliography{IEEEabrv,./bibliography}

\section*{Appendix}
First, let us state some quantities and principles useful for all proofs. Let $\vec{F} = (\vec{\imath}_0,\vec{\jmath}_0)F$, $\vec{F}_a = (\vec{\imath}_0,\vec{\jmath}_0)F_a$, 
$\vec{g} = (\vec{\imath}_0,\vec{\jmath}_0)ge_1$, $\vec{\jmath} = (\vec{\imath}_0,\vec{\jmath}_0)Re_2$, 
$\vec{v}_w~=~(\vec{\imath}_0,\vec{\jmath}_0)\dot x_w$,
$\vec{v}_r~=~(\vec{\imath}_0,\vec{\jmath}_0)\dot x_r$, $\vec{a}_r = (\vec{\imath}_0,\vec{\jmath}_0)\ddot x_r$, and 
$\vec{v}_a = (\vec{\imath}_0,\vec{\jmath}_0)\dot x_a$.

Then, note that the existence of an equilibrium orientation such that \eqref{equilibiumOrinetationCo} 
holds is equivalent to the existence, at any fixed time $t$, of one zero of the following function
\begin{IEEEeqnarray}{RCL}
   f_t(\theta) &:=& F^T(\dot{x}_r(t),\theta,t)R(\theta)e_2,
  \label{eq:wnugeneralp} 
\end{IEEEeqnarray}
where  
\begin{IEEEeqnarray}{RCL}  
   \IEEEyesnumber
    \label{FProofEq}
   F(\dot{x},\theta,t) &=& F_{gr}(t) + F_a(\dot{x}_a,\alpha(\dot{x}_a,\theta)), \IEEEyessubnumber \label{FwithFgrFa} \\
   F_{gr} &:=& mge_1-m \ddot x_r, \IEEEyessubnumber \label{PFgr} \\
   F_a &=& k_a |\dot{x}_a|[c_L(\alpha)S-c_D(\alpha)I]\dot{x}_a, \IEEEyessubnumber \label{PFa} \\
   \dot x_a &=& \dot x - \dot x_w, \IEEEyessubnumber \label{PxaD} \\
   \alpha  &=& \theta -\gamma  +(\pi -\delta) \IEEEyessubnumber \label{Pa} \\
   \gamma &=& \text{atan2}(\dot x_{a_2},\dot x_{a_1}). \IEEEyessubnumber \label{Pg}
\end{IEEEeqnarray}

\section*{Proof of Lemma~\ref{lemma:dissipativityNotSufficient}}
\label{Plemma:dissipativityNotSufficient}

In  coordinates, the aerodynamic force passivity \eqref{dissipativity} 
writes 
\begin{IEEEeqnarray}{r}
  \dot x^T_a F_a \leq 0 \quad \forall (\dot x_a,\alpha). \IEEEeqnarraynumspace \label{Pdissipativity}
\end{IEEEeqnarray}
To show that \eqref{Pdissipativity} does not in general imply the existence of an equilibrium orientation,
it  suffices to find an aerodynamic force satisfying \eqref{Pdissipativity}  such that 
\eqref{eq:wnugeneralp} never crosses zero for some reference and wind velocities at a time instant. Choose
\begin{equation}
      \label{PcoefficientsPassNoEqui}
	\begin{system}
	  c_{L}(\alpha) &=&  \sin(\alpha) \\
	  c_{D}(\alpha) &=& c_{0} + 1 - \cos(\alpha) > 0, \quad \forall \alpha,
	\end{system}
\end{equation}
with $c_{0} > 0$. It is then straightforward to verify that the aerodynamic force given by \eqref{PFa} with 
the coefficients \eqref{PcoefficientsPassNoEqui} satisfies~\eqref{Pdissipativity}; in addition, note also that 
$c_{L}(0) = c_{L}(\pi) = 0.$
Since the vector $F$ on the
right hand side of~\eqref{eq:wnugeneralp} is evaluated at the reference velocity,
we have to evaluate the quantities \eqref{FProofEq} at  $\dot x_r$. Let us  assume that

{\bf{A1:}} the thrust force is perpendicular to the zero lift direction so that $\delta = \pi/2$;

{\bf{A2:}} there exists a time $\bar t$ such that
 \begin{enumerate}
  \item[i)]  the reference and wind velocities imply 
  $\gamma(\dot x_r(\bar t) {-} \dot x_w(\bar t)){=} \pi/2 
  \ \ \text{ and } \ \ k_a|\dot x_r(\bar t) {-}\dot x_w(\bar t)|^2 {=} 1;   $
  \item[ii)]  the reference acceleration $\ddot x_r(\bar t)$ implies \\
  $F_{gr_1}(\bar t) = 0  \ \ \text{ and } \ \ F_{gr_2}(\bar t) = c_0 + 1. $
 \end{enumerate}
 By evaluating the angle of attack \eqref{Pa}  at the reference velocity with 
 $\bold{A1}$ and $\bold{A2}$i,
 one  verifies that $\alpha(\bar t) = \theta$. Then, \eqref{eq:wnugeneralp} at $t = \bar t$ becomes 
   $f_{\bar t}(\theta) = [F_{gr_2}(\bar t) {-}c_D(\theta)]\cos(\theta) + [c_L(\theta){-}F_{gr_1}(\bar t)]\sin(\theta). $
In view of the aerodynamic coefficients \eqref{PcoefficientsPassNoEqui} and $\bold{A2}$ii, one has
$f_{\bar t}(\theta) \equiv 1 \ne 0. $
Hence, there exists an aerodynamic force
that satisfies \eqref{Pdissipativity} but for which there does not exist an equilibrium orientation.


\section*{Proof of Theorem~\ref{th:existence}}
\label{proof:thExistence}
\subsection*{Proof of the item i)}
Assume that the thrust force is parallel to the zero-lift-line so that $\delta = 0$.
The existence of the equilibrium orientation for $\delta = \pi$ can be proven using the same arguments as those below.
Now, in view of Eqs.~\eqref{eq:VaComponents}, 
$\dot x_a = R(\theta) v_a$, $S = R^T(\theta)SR(\theta) $, and of $\delta = 0$, one  verifies that 
\eqref{eq:wnugeneralp} is
\begin{IEEEeqnarray}{RCL}
  \label{fThetaSym}
   f_t(\theta) &=&  F^T_{gr}(t)R(\theta)e_2 \nonumber \\
   &-& k_a|\dot x_{rw}(t)|^2[c_L(\alpha_r)\cos(\alpha_r)+c_D(\alpha_r)\sin(\alpha_r)], \nonumber 
\end{IEEEeqnarray}
where $F_{gr}(t)$ is given by \eqref{PFgr} and 
\begin{IEEEeqnarray}{RCL}
  \label{parametersR}
   \alpha_r(\theta,t)  &=& \theta -\gamma_r(t)  +\pi   \label{alphar}  ,\\
   \IEEEyesnumber
   \gamma_r(t) &=& \text{atan2}(\dot x_{rw_2},\dot x_{rw_1}) \IEEEyessubnumber \label{Pgr},\\
   \dot x_{rw}(t) &:=& \dot x_{r}(t) - \dot x_{w}(t).\IEEEyessubnumber \label{xrw}
\end{IEEEeqnarray}
It follows from \eqref{alphar} that at any time $t$ there exists an orientation $\theta_0(t)$ such that 
$\theta = \theta_0(t)$ yields $\alpha_r(t) = 0$, i.e.
\[\theta = \theta_0(t) =\gamma_r(t)-\pi \quad \Rightarrow \quad \alpha_r(t) = 0.\]
Then, $\theta = \theta_0(t) + \pi$ yields $\alpha_r(t) = \pi$ and 
$\theta = \theta_0(t) - \pi$ yields $\alpha_r(t) = -\pi$. 
Since it is assumed that the body shape is symmetric, then~\eqref{cLZeroCLPiUZ} holds.
Thus, 
  Eq. \eqref{fThetaSym} yields
\begin{IEEEeqnarray}{RCL}
 \label{propertiesfSym}
 f_t(\theta_0(t) + \pi) &=& f_t(\theta_0(t) - \pi) = -f_t(\theta_0(t))
\end{IEEEeqnarray}
because $e^T_2R^T(\theta_0+\pi)F_{gr}(t)=e^T_2R^T(\theta_0-\pi)F_{gr}(t) = -e^T_2R^T(\theta_0)F_{gr}(t)$.
In view of \eqref{propertiesfSym}, the proof of the existence of (at least) two zeros of the function $f_t(\theta)$ at any fixed time $t$, and thus of
two equilibrium orientations,
is then a direct application of the \emph{intermediate value theorem}. In fact, by assumption, $f_t(\theta)$ is continuous versus $\theta$ ($c_L$ and $c_D$ are continuous) 
and defined $\forall t$ ($\dot{x}_r$ is differentiable). These two zeros, denoted by $\theta_{e_1}(t)$ and $\theta_{e_2}(t)$, 
belong to $\theta_{e_1}(t) \in [\theta_0(t)-\pi, \theta_0(t)]$ and $\theta_{e_2}(t) \in [\theta_0(t), \theta_0(t) + \pi]$. 

\begin{remark}
The key assumption for the above is  
 $c_L(0) = c_L(\pi) = 0$. 
Hence, drag forces have no role in the existence of an equilibrium orientation of  symmetric shapes 
with a
thrust force parallel to their axis of symmetry. 
If the thrust force is not parallel to the axis of symmetry, 
one  shows that $c_L(0) = c_L(\pi) = 0$ is no longer sufficient to ensure the equilibrium orientation existence\footnote{Use the same counterexample used to prove Lemma \ref{lemma:dissipativityNotSufficient}.}. 
\end{remark}
\subsection*{Proof of the item ii)}
Under the assumption of bisymmetric bodyshape, Eqs.~\eqref{LiftPropertiesBSymmetricB} hold, i.e.
	$c_D( \alpha) = c_D( \alpha {\pm} \pi) \ \forall \alpha$, 
	$c_L( \alpha) = c_L( \alpha {\pm} \pi) \ \forall \alpha.$
	In view of~\eqref{PFa}, this aerodynamic property implies
$F_a(\dot{x}_a,\alpha) = F_a(\dot{x}_a,\alpha {\pm} \pi). $ 
Consequently, using the expression
of the angle of attack in~\eqref{Pa}, one verifies that the apparent external force given by \eqref{FwithFgrFa} satisfies
\begin{IEEEeqnarray}{RCL}
  \label{PropFBisymmProof}
  F(\dot{x},\theta,t) = F(\dot{x},\theta \pm \pi,t)  \quad \forall (\dot x,\theta, t). 
\end{IEEEeqnarray} 
In turn, it is straightforward to verify that the function $f_t(\theta)$ given by~\eqref{eq:wnugeneralp} satisfies, at any time $t$, the following 
\begin{IEEEeqnarray}{RCL}
  \label{PropfBisymm2}
  f_t(\theta + \pi) = f_t(\theta - \pi) =-f_t(\theta) \quad \forall \theta. 
\end{IEEEeqnarray} 
Then, analogously to the proof of the  Item 1), the existence of at least two equilibrium orientations  $\theta_{e_1}(t)$ and $\theta_{e_2}(t)$
such that $f_t(\theta_{e_1}(t))= f_t(\theta_{e_2}(t)) = 0$
can be shown by applying the \emph{intermediate value theorem}. 

Observe that Eqs. \eqref{PropfBisymm2} imply that
if $\theta_{e_1}(t)$ is an equilibrium orientation, i.e. $f_t(\theta_{e_1}(t)) = 0 \quad \forall t$, then another equilibrium orientation is given by $\theta_{e_2}(t) = \theta_{e_1}(t) + \pi$.
Now, to show that there always exists an equilibrium orientation ensuring a positive-semi definite thrust intensity, 
from Eq.~\eqref{equilibiumThrustCondition}  observe that the thrust intensity at the equilibrium point is given by
   $T_e = F^T(\dot{x}_r(t),\theta_e(t),t)R(\theta_e(t))e_1.$
Then, it follows from \eqref{PropFBisymmProof} that if the thrust intensity is negative-semi definite at $t$ along an equilibrium orientation, i.e. 
$T_e(\dot{x}_r(t),\theta_{e_1}(t),t) \leq 0$, then it is positive-semi definite at 
$\theta_{e_2}(t){=}\theta_{e_1}(t){+}\pi$, i.e. $T_e(\dot{x}_r(t),\theta_{e_1}(t)+\pi,t) \geq 0.$ Hence, one can always build up an equilibrium orientation
$\theta_e(t)$
associated with a positive-semi definite  thrust intensity. 
%

\section*{Proof of Theorem~\ref{th:existenceDeltaDifZeroSymm}}
\label{proof:thExistenceMune0}
First, observe that
if $\sin(\delta) = 0$, then the equilibrium orientation existence follows 
from Theorem \ref{th:existence} since the thrust force is parallel to the zero-lift-direction. Hence assume that 
\begin{IEEEeqnarray}{RCL}
  \label{sinDeltaDiff0}
  \sin(\delta) \ne 0.
\end{IEEEeqnarray} 
Recall that  the equilibrium orientation existence 
is equivalent to the existence, at any fixed time $t$, of one zero of the function $f_t(\theta)$ given by 
\eqref{eq:wnugeneralp}. In view of \eqref{eq:VaComponents}, $\dot x_a = R(\theta) v_a$, and of $S =  R^T(\theta)SR(\theta)$, 
one can verify that \eqref{eq:wnugeneralp} becomes
\begin{IEEEeqnarray}{RCL}
  \label{fThetaBSym}
   f_t(\theta) =  F^T_{gr}(t)R(\theta)e_2 &-& k_a|\dot x_{rw}(t)|^2[c_L(\alpha_r)\cos(\alpha_r+\delta) \nonumber \\
                                         &+&\quad c_D(\alpha_r)\sin(\alpha_r+\delta)], 
  \IEEEeqnarraynumspace
\end{IEEEeqnarray}
where $F_{gr}$ is given by \eqref{PFgr}, 
\begin{IEEEeqnarray}{RCL}
   \alpha_r =\alpha_r(\theta,t) = \theta -\gamma_r(t)  +\pi -\delta, \label{alphar2}
\end{IEEEeqnarray}
$\gamma_r$ by \eqref{Pgr},
and $\dot x_{rw}$  by \eqref{xrw}. 
From Eq. \eqref{fThetaBSym} 
note that if $|\dot x_{rw}(t)| = 0$, 
then there exist at least two zeros for the function $f_t(\theta)$, i.e. at least two equilibrium orientations at the time~$t$. Thus, let us  focus on the following case  
\begin{IEEEeqnarray}{RCL}
  \label{relRefWindDiff0}
  |\dot x_{rw}(t)| \ne 0.
\end{IEEEeqnarray} 
It follows from \eqref{alphar2} that at any fixed time $t$,
there exists an orientation $\theta_0(t) = \gamma_r(t)-\pi+\delta$ such that 
$\theta = \theta_0(t)$ yields $\alpha_r = 0$, so
 $\theta = \theta_0(t) + \pi$ yields $\alpha_r = \pi$. Now, if 
 $f_t(\theta_0(t))f_t(\theta_0(t)+ \pi) \leq 0,$
then there exists a zero for the function $f_t(\theta)$, and this zero belongs to $[\theta_0(t),\theta_0(t)+ \pi]$: in fact, the function $f_t(\theta)$ changes sign on this domain and is continuous versus $\theta$. We are thus interested in the case when the above inequality is not satisfied. 
Hence, 
assume also that
\begin{IEEEeqnarray}{RCL}
  \label{noEquilibriumYet}
  f_t(\theta_0(t))f_t(\theta_0(t)+\pi) > 0.
\end{IEEEeqnarray} 
Given the assumption that the body's shape is symmetric, one has $c_L(0)=c_L(\pi) = 0$. So, in view of \eqref{fThetaBSym}, imposing
\eqref{noEquilibriumYet} divided by $k^2_a|\dot{x}_{rw}(t)|^4\sin(\delta)^2$, which we recall to be assumed different from zero, yields  
\begin{IEEEeqnarray}{RCL}
  \label{noEquilibriumYet1}
  [a_t-c_D(0)][c_D(\pi)-a_t] > 0,
\end{IEEEeqnarray}  
where
  $a_t := \tfrac{F^T_{gr}(t)R(\theta_0(t))e_2}{k_a\sin(\delta)|\dot{x}_{rw}(t)|^2}$. 
Under the assumption that $c_D(0) < c_D(\pi)$,
the inequality \eqref{noEquilibriumYet1} implies that
\begin{IEEEeqnarray}{RCL}
  c_D(0) <a_t< c_D(\pi).   \label{ContrstaintNoEquilibrium}
\end{IEEEeqnarray}  
When the constraint \eqref{ContrstaintNoEquilibrium} is satisfied, the inequality \eqref{noEquilibriumYet} holds and we cannot (yet)
claim the existence of an equilibrium orientation at the time instant $t$. The following shows that when the inequality \eqref{ContrstaintNoEquilibrium} 
is satisfied, the existence of an equilibrium orientation at the time instant $t$ follows from the symmetry of the body's shape provided that the
conditions of Theorem~\ref{th:existenceDeltaDifZeroSymm} hold true. 
Recall that when the inequality \eqref{ContrstaintNoEquilibrium}  is not satisfied, the existence of an equilibrium orientation at the time $t$ follows from the 
fact that 
$f_t(\theta_0(t))f_t(\theta_0(t)+ \pi) \leq 0$.

Now,  under the assumption
that the body's shape is symmetric, we have that Eqs.~\eqref{LiftPropertiesSymmetricB} hold.
Let $\bar \alpha \in \mathbb{R}^+$; then, by using 
    $c_D( \alpha) = c_D( -\alpha)$, $c_L( \alpha)~=~-c_L( -\alpha),$
and \eqref{fThetaBSym},
one verifies that (recall that $\theta = \theta_0(t) \Rightarrow \alpha _r(t) = 0$, so $\theta = \theta_0(t) \pm \bar \alpha \Rightarrow \alpha _r(t) = \pm \bar \alpha$):
\begin{IEEEeqnarray}{RCL}
  \label{noEquilibriumYet2}
  f_t(\theta_0(t)&-& \bar \alpha)f_t(\theta_0(t) + \bar \alpha ) 
             =\Delta_a^2 \sin^2(\delta)-\Lambda_b^2,
  \IEEEeqnarraynumspace  \\
  \label{DeltaLambda}
  \Delta_a=\Delta_a(\bar \alpha)  &:=& [a_t - c_D(\bar \alpha)]\cos(\bar \alpha) + c_L(\bar \alpha)\sin(\bar \alpha), \IEEEeqnarraynumspace  \label{Delta}
\end{IEEEeqnarray} 
$\Lambda_b(\bar \alpha) := [b_t + c_D(\bar \alpha)\cos(\delta)]\sin(\bar \alpha)+ c_L(\bar \alpha)\cos(\bar \alpha)\cos(\delta)$,   $b_t := \tfrac{F^T_{gr}(t)R(\theta_0(t))e_1}{k_a|\dot{x}_{rw}(t)|^2}$.
It follows from Eq. \eqref{noEquilibriumYet2} that if 
\begin{IEEEeqnarray}{RCL}
  \label{Delta1}
  \forall a_t  : c_D(0)  <a_t< c_D(\pi), \exists \bar \alpha_a \in \mathbb{R} \ : \ \Delta_a(\bar \alpha_a) = 0, \IEEEeqnarraynumspace
\end{IEEEeqnarray} 
then there exists a zero for the function $f_t(\theta)$, and this zero belongs to  $[\theta_0(t){-}\bar \alpha_a,\theta_0(t){+}\bar \alpha_a]$ (the function $f_t(\theta)$ would change sign in this domain).
The existence of an $\bar{\alpha}_a$ such that \eqref{Delta1} holds can be deduced by imposing that 
\begin{IEEEeqnarray}{RCL}
  \label{Delta2}
  \forall a_t \ :\ && c_D(0)  <a_t< c_D(\pi), \quad \exists \alpha_0, 
  \alpha_s \in \mathbb{R}, 
  \ : 
  \nonumber \\&& 
  \Delta_a(\alpha_0)\Delta_a(\alpha_s) \leq 0, \IEEEeqnarraynumspace
\end{IEEEeqnarray} 
which implies \eqref{Delta1} with $\bar \alpha_a \in [\alpha_0,\alpha_s]$ since $\Delta_a(\bar \alpha)$ is continuous versus $\bar \alpha$. 
Now, 
in view of \eqref{Delta} note that $\forall a_t : c_D(0)  <a_t< c_D(\pi)$ one has
  $\Delta_a(0) > 0 $.
Still from \eqref{Delta}, note also that $\forall a_t  : c_D(0)  {<}a_t{<}~c_D(\pi)$ one has
\begin{IEEEeqnarray}{RCL}
  \Delta_a(\bar \alpha)  &\leq& c_D(\pi)|\cos(\bar \alpha)| - c_D(\bar \alpha)\cos(\bar \alpha) + c_L(\bar \alpha)\sin(\bar \alpha).\nonumber 
\end{IEEEeqnarray} 
If there exists  $\alpha_s \in (0,90^\circ)$ such that $c_L(\alpha_s) > 0$ and 
\begin{IEEEeqnarray}{RCL}
   \text{Cond.~\eqref{conditionMuNe0}}  {\ \Leftrightarrow \ } c_L(\alpha_s)\sin(\alpha_s){-}[c_D(\alpha_s){-}c_D(\pi)]\cos(\alpha_s) {\leq} 0  \nonumber \nonumber \IEEEeqnarraynumspace
\end{IEEEeqnarray} 
then $\Delta_a(\alpha_s) \leq 0$ and \eqref{Delta2} holds with $\alpha_0 = 0$. Consequently, there exists an angle $\bar{\alpha}_a$
such that \eqref{Delta1} is satisfied and, subsequently, an equilibrium orientation $\theta_e(t)$.

\end{document}